\documentclass[aps,pre,showpacs,floatfix]{revtex4}
\usepackage{epsf}
\usepackage{subfigure}
\usepackage{amsmath}
\usepackage{amssymb}
\usepackage{graphicx}
\usepackage{epstopdf}
\newcommand{\nc}{\newcommand}
\nc{\beq}{\begin{equation}}
\nc{\eeq}{\end{equation}}
\nc{\bega}{\begin{eqnarray}}
\nc{\ega}{\end{eqnarray}}
\nc{\6}{\partial}
\nc{\non}{\nonumber}

\nc{\Dre}{D^{(r)}} %%don't works with \nc{Dre}{D^{(r)}}
\nc{\rhoA}{\rho_A}
\nc{\rhoB}{\rho_B}
\nc{\rhoav}{\bar{\rho}}
\nc{\rhoplus}{\rho_+}
\nc{\rhominus}{\rho_-}
\nc{\dd}{{\rm d}}
\nc{\fA}{f_A}
\nc{\fB}{f_B}
\nc{\jA}{\vec{j}_A}
\nc{\jB}{\vec{j}_B}
\nc{\fonere}{f_1^{(\rm Re)}}
\nc{\foneim}{f_1^{(\rm Im)}}

\begin{document}

%
%\date{\today}
%
\title{%
Entropy production in diffusion-reaction systems: 
The reactive random Lorentz gas}
\author{L\'aszl\'o M\'aty\'as and Pierre Gaspard} 
\affiliation{Center for Nonlinear Phenomena and Complex Systems,\\
Universit\'e Libre de Bruxelles, Code Postal 231, Campus Plaine, 
B-1050 Brussels, Belgium}
%
%\maketitle

\begin{abstract}
We report the study of a random Lorentz gas with a reaction of
isomerization $A\rightleftharpoons B$ between the two colors
of moving particles elastically bouncing on hard disks.
The reaction occurs when the moving particles collide
on catalytic disks which constitute a fraction of all the disks.
Under the dilute-gas conditions, the reaction-diffusion process 
is ruled by two coupled Boltzmann-Lorentz equations
for the distribution functions of the colors.  The
macroscopic reaction-diffusion equations with cross-diffusion
terms induced by the chemical reaction are derived from the
kinetic equations. We use a $H$-theorem of the kinetic theory in order
to derive a macroscopic entropy depending on the gradients
of color densities and which has a non-negative entropy production
in agreement with the second law of thermodynamics.
\end{abstract}

%\draft

\maketitle

%%***************************************************************
%%***************************************************************
\section{Introduction}

During the last decades, irreversibility and transport properties have been intensively
studied in low-dimensional dynamical systems
\cite{EvHobooks,GaNi90,Ga92,ChEyLeSi93,Ru96,Gabook,Dobook,%
VoTeBr98,TeVo00,MaTeVo00,TaGa00,TeVoMa01,GiDo99,GiDoGa00,DoGaGi02,GaNiDo02}. 
It has been shown that the macroscopic transport coefficients 
can be related to the characteristic quantities of the microscopic dynamics 
\cite{GaNi90,Gabook}. 
Moreover, results have also been obtained for the entropy balance in
dynamical systems sustaining transport processes of diffusion \cite{GiDoGa00,DoGaGi02,GaNiDo02}, 
electric conduction \cite{VoTeBr98}, cross effects 
\cite{MaTeVo00,TaGa00}, and shear viscosity \cite{TeVoMa01,ViGa03a,ViGa03b}.
For these processes, the irreversible entropy production of nonequilibrium
thermodynamics could be derived from the underlying microscopic dynamics.

Beside transport, reaction processes have also been investigated from the
viewpoint of dynamical systems theory.  Reaction-diffusion processes play a
crucial role in physico-chemical systems far from the thermodynamic
equilibrium and they have been much studied at the
macroscopic level of description on the basis of the phenomenological
nonequilibrium thermodynamics \cite{NiPr77}.

Recently, several microscopic models of reaction-diffusion processes
have been introduced and analyzed in order to understand the
foundations of the phenomenological assumptions.
Multibaker and Lorentz gas models of reaction-diffusion processes
were introduced in Refs. \cite{GaKl98,Ga99,footnote1}.  In this early version,
the reaction was an isomerization $A\rightleftharpoons B$
with unit probability upon the passage of the particle on a catalytic side where
the reaction occurs.  In Ref. \cite{ClGa00}, a spatially periodic reactive Lorentz gas was
introduced in which the isomerization $A\rightleftharpoons B$ occurs
with a probability $0\leq p_0\leq 1$ when the particle collides on catalytic disks.
The catalytic disks are few among the disks which compose the
Lorentz gas. Spatially periodic Lorentz gas and multibaker models
with reaction $A\rightleftharpoons B$ have been studied in detail
and their diffusive and reactive modes were constructed together
with their dispersion relations \cite{ClGa00,ClGa02,GaCl02}.  In this way, the macroscopic
reaction-diffusion equations could be derived in a systematic way
from the underlying dynamics.  The analysis revealed the existence
of cross-diffusion effects induced by the reaction.  Such cross diffusion
is typically overlooked in the phenomenological approach \cite{dGM,KoPr98}.  Moreover,
the entropy production of the standard phenomenological entropy may be
negative because of the cross-diffusion effects.  Although this problem
only happens for extreme particle densities it nevertheless sheds some doubts
on the phenomenological assumptions.

The purpose of the present paper is to clarify these issues and understand the
problem of entropy production in such reaction-diffusion systems.  For this purpose,
we consider a {\it random} Lorentz gas with a fraction of catalytic disks where the
isomerization $A\rightleftharpoons B$ occurs with a given probability $p_0$.
The color $A$ or $B$ carried by the moving particle may correspond
to the spin of the particle, in which case the catalytic disks model some
spin-flipping impurities in the system.  For dilute Lorentz gases, we can use kinetic
theory and linear Boltzmann-Lorentz equations \cite{Be82,BoBuSi83}.
Thanks to such master equations, we can derive a $H$-theorem which allows us
to obtain an expression for the entropy. We prove that the
corresponding entropy production is non-negative 
with respect to the time evolution induced by the 
macroscopic reaction-diffusion equations.

The plan of the paper is the following.
The kinetic equations of the reactive Lorentz gas are introduced in Sec. 
\ref{sec:Lorentzgas} where we prove the $H$-theorem.
The macroscopic reaction-diffusion equations are derived in Sec. 
\ref{sec:solutions}.  The entropy balance is obtained and discussed in
Sec. \ref{sec:entropybalance}.  Conclusions are drawn in Sec.
\ref{sec:conclusions}.

%%*******************************************************************
%%*******************************************************************
\section{The reactive Lorentz gas}
\label{sec:Lorentzgas}
  
In the random Lorentz gas, fixed circular scatterers correspond
to heavy particles and moving point particles to the light ones. 
The system is defined by the density of the scatterers $n$, 
their radius $a$, and by the velocity of the light particles 
$\vec{v}=v(\cos\varphi,\sin\varphi)$, the system being 
two dimensional. 
Because the magnitude of the velocities is fixed, 
the velocity vector can be characterized solely by the angle $\varphi$.  
Here, we consider a random distribution of the disks with 
low density $n \ll a^{-2}$.

The reactive Lorentz gas \cite{ClGa00} consists of 
two different types of light particles $A$ and $B$, that have a free flight 
between the collisions. 
Some of the fixed scatterers act as catalysts, 
i.e., if an $A$ particle collides with such a scatterer 
it becomes a $B$ with the probability $p_0$, and vice versa. 
The catalysts have a density $n_r$. 
The reaction scheme is
\bega
A + {\rm disk} \, & \rightleftharpoons &  A + {\rm disk}\; ,  \\
B + {\rm disk} \, & \rightleftharpoons &  B + {\rm disk} \; , \\
A + {\rm catalyst} \, & \rightleftharpoons &  B + {\rm catalyst} \; , \qquad \mbox{with probability} \ p_0\; .
\ega
The colors can be considered as the two components of a spin
one-half carried by the moving particles.
The evolution of the system can be characterized by the 
two distribution functions of the two components 
$\fA(\vec{r},\vec{v})$ and $\fB(\vec{r},\vec{v})$. 
The integral over the velocities of these functions 
gives the number density at point $\vec{r}$ of each component $A$ or $B$. 
Having in view that the time evolution of the 
distribution function 
$\fA$ (resp. $\fB$) is also influenced by the presence of 
the other component $B$ (resp. $A$) that may collide with a catalyst, we obtain 
the system of equations  
\bega
\6_t \fA + \vec{v} \cdot \vec{\nabla} \fA
&=&  \frac{av (n-p_0 n_r) }{2} 
\int_{-\pi}^{+\pi} d\varphi' \left| \sin \frac{\varphi-\varphi'}{2} \right| 
[\fA(\vec{r},\varphi') - \fA(\vec{r}, \varphi)]  \non \\
&&+ \frac{av p_0 n_r}{2} 
\int_{-\pi}^{+\pi} d\varphi' \left| \sin \frac{\varphi-\varphi'}{2} \right| 
[\fB(\vec{r},\varphi') - \fA(\vec{r}, \varphi)]\;  , 
\label{eq:compA}  \\
\6_t \fB + \vec{v} \cdot \vec{\nabla} \fB 
&=&  \frac{av (n-p_0 n_r) }{2} 
\int_{-\pi}^{+\pi} d\varphi' \left| \sin \frac{\varphi-\varphi'}{2} \right| 
[\fB (\vec{r},\varphi') - \fB (\vec{r}, \varphi)]  \non \\
&&+ \frac{av p_0 n_r}{2} 
\int_{-\pi}^{+\pi} d\varphi' \left| \sin \frac{\varphi-\varphi'}{2} \right| 
[\fA (\vec{r},\varphi') - \fB(\vec{r}, \varphi)] \; . 
\label{eq:compB}
\ega 

One can observe that the total system of the moving particles
$A$ and $B$ follows a diffusive process 
that can be characterized by the the sum 
of the distribution functions
\beq
f\equiv \fA + \fB\; .
\eeq
The time evolution of $f$ is given by a linear Boltzmann equation, 
also known as the Boltzmann-Lorentz equation \cite{Be82,BoBuSi83}: 
\beq
\label{eq:ftotal}
\6_t f + \vec{v} \cdot \vec{\nabla} f 
= 
\frac{a v n}{2} 
\int_{-\pi}^{+\pi} d\varphi' \left| \sin \frac{\varphi-\varphi'}{2} \right| 
[f(\vec{r},\varphi') - f(\vec{r}, \varphi)] \; .
\eeq
This equation rules the so-called {\em diffusion sector}.
In order to have the time evolution of both distribution functions, 
we need a further equation beside Eq.~(\ref{eq:ftotal}). 

If we introduce the quantity
\beq
g\equiv \fA - \fB\; ,
\eeq
and take the difference of Eqs. (\ref{eq:compA}) and (\ref{eq:compB}), 
we obtain the equation
\beq 
\6_t g + \vec{v} \cdot \vec{\nabla} g = \frac{av (n - 2p_0 n_r)}{2} 
\int_{-\pi}^{+\pi} d \varphi' 
\left| \sin \frac{\varphi-\varphi'}{2} \right| 
[g(\vec{r},\varphi') -  g(\vec{r},\varphi)]   
- 4 av p_0 n_r g(\vec{r},\varphi)\; ,
\label{eq:gdiff}
\eeq 
which rules the {\em reaction sector}.
One can notice that the last equation describes a decay in time 
of the function $g$ that can be separated from the rest of the solution 
\beq
g (\vec{r}, \varphi, t) = {\rm e}^{-4av p_0 n_r \, t} h (\vec{r}, \varphi) \; .  
\eeq 
The equation for $h$ reads 
\beq
\label{eq:hdiff}
\6_t h +\vec{v} \cdot \vec{\nabla} h 
= 
\frac{av(n - 2p_0 n_r)}{2} 
\int_{-\pi}^{+\pi} d\varphi' \left| \sin \frac{\varphi-\varphi'}{2} \right| 
[h(\vec{r},\varphi') 
-  h(\vec{r},\varphi) ]  \; ,
\eeq 
which has the same form as Eq.~(\ref{eq:ftotal}). 

%%**************************************************************
%%**************************************************************
\subsection{$H$-theorem}
\label{sec:H-theorem}

The Boltzmann entropy of the system of Eqs. (\ref{eq:compA}) and (\ref{eq:compB}) is given by the sum 
of the entropies of the two components:
\beq
s=  \int  d \varphi \; \left( \fA \ln \frac{f^0}{\fA} + \fB \ln \frac{f^0}{\fB} \right)\; ,
\label{eq:s-Boltzmann}
\eeq
where Boltzmann's constant is taken equal to unity, $k_{\rm B}=1$.

Inserting to the integral on the r.h.s. the time evolution Eqs.
(\ref{eq:compA}) and (\ref{eq:compB}),   
we get for the variation of the entropy  
\beq
\6_t s 
=  \int d \varphi 
        \left\{ \ln \frac{f^0} {{\rm e}\fA}
                \left[-\vec{v} \cdot\vec{\nabla} \fA + C_A (\fA,\fB)\right] 
               + \ln \frac{f^0} {{\rm e}\fB}
                \left[-\vec{v} \cdot\vec{\nabla} \fB + C_B (\fA,\fB)\right]  
         \right\} \; ,
\eeq
where the collision integral $C_A (\fA,\fB)$ has the form  
\beq 
C_A (\fA,\fB)=  
\int_{-\pi}^{+\pi} d \varphi'
\left| \sin \frac{\varphi-\varphi'}{2} \right| 
\left[
\frac{av (n-p_0 n_r) }{2} 
(\fA'- \fA)  
+ \frac{av p_0 n_r}{2}  
(\fB' - \fA) \right] \; ,
\eeq 
with the notations:
\bega
\fA &=& \fA(\vec{r},\varphi,t) \; ,\\
\fA' &=& \fA(\vec{r},\varphi',t) \; ,\\
\fB &=& \fB(\vec{r},\varphi,t) \; ,\\
\fB' &=& \fB(\vec{r},\varphi',t) \; .
\ega

This relation can be written in form of balance \cite{dGM} 
\beq
\6_t s = - \vec{\nabla} \cdot \vec{J}_s + \sigma_s \; ,
\eeq
where
\beq
\label{eq:sBoltz-flow}
\vec{J}_s = 
     \int d \varphi \; \vec{v} \; 
\bigg( \fA \ln \frac{f^0}{\fA} + \fB \ln \frac{f^0}{\fB}
             \bigg) 
\eeq 
is the entropy current and
\bega 
\sigma_s 
= \frac{av}{4} 
        \int  d \varphi d\varphi'  
           \left| \sin \frac{\varphi-\varphi'}{2} \right|           
    \bigg\{  (n-p_0 n_r) 
   & \left[ (\fA'-\fA) 
    \ln \frac{\fA'}{\fA} 
    +  (\fB'-\fB) 
    \ln \frac{\fB'}{\fB} 
    \right] &\non \\
+ p_0 n_r  &\left[ 
    (\fB'-\fA) 
    \ln \frac{\fB'}{\fA} 
    +  (\fA'-\fB) 
    \ln \frac{\fA'}{\fB} 
    \right]& \bigg\} \geq 0
\label{eq:entrgen}
\ega
is the entropy production. One can see on this general form of the entropy production 
that it is positive definite under the consistency condition that
\beq
0 \leq p_0 n_r \leq n \; .
\label{consistency}
\eeq

%%**************************************************************
%%**************************************************************
\section{Derivation of the macroscopic reaction-diffusion equations}
\label{sec:solutions}

\begin{figure}[htbp]
\centerline{\includegraphics[width=10cm]{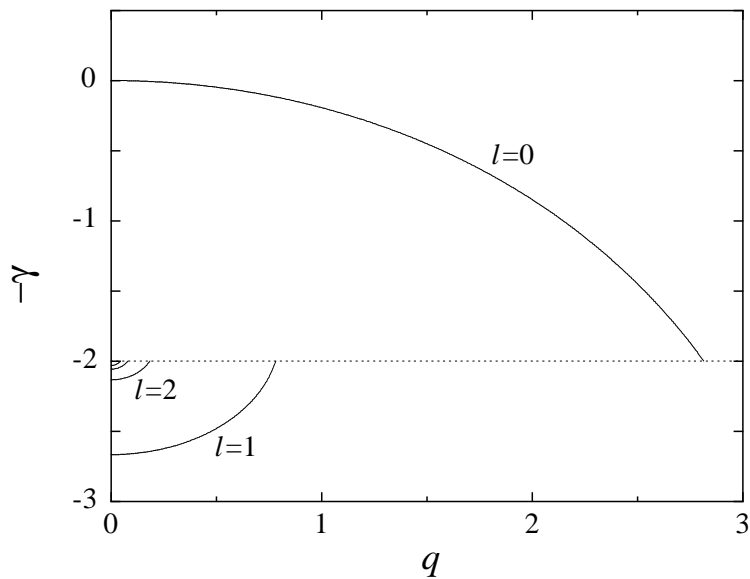}}
\caption{Eigenvalues $-\gamma$ of Eq. (\ref{eigen.eq}) versus the wavenumber $q=q_x$ for $q_y=0$
and $c=1$.}
\label{fig1}
\end{figure}

A convenient way to find the solution of 
Eqs.\ (\ref{eq:ftotal}) and (\ref{eq:hdiff})
\beq
\label{eq:X}
\6_t X + \vec{v} \cdot \vec{\nabla} X= \frac{c}{2}
\int_{-\pi}^{+\pi} d\varphi' \left| \sin \frac{\varphi-\varphi'}{2} \right| 
[X(\vec{r},\varphi') - X(\vec{r}, \varphi)] \; ,
\eeq
 is by writing 
the distribution functions $X=f$ or $h$ as Fourier series in the velocity 
angle $\varphi$
\beq
X(\vec{r}, \varphi,t) = \sum_{l=-\infty}^{+\infty} X_l(\vec{r}, t) \, {\rm e}^{il \varphi} \; .
\label{eq:XFourier}
\eeq
As a consequence of Eqs. (\ref{eq:ftotal}) or (\ref{eq:hdiff}), the 
Fourier components $X_l$ fulfill the following coupled 
differential equations
\beq
\6_t X_l +v \frac{\6_x -i \6_y}{2} X_{l-1} 
         +v \frac{\6_x +i \6_y}{2} X_{l+1} 
= \frac{8 l^2c}{1-4 l^2} X_l  \; ,
\label{eq:PDEFourier}
\eeq
with 
\beq
c=avn \qquad \mbox{for}\quad X=f \; ,
\eeq 
and
\beq
c=av(n-2p_0n_r) \qquad \mbox{for}\quad X=h \; .
\eeq 
The distribution function $X$ being a real function, the Fourier coefficients 
with negative index are the complex conjugates 
of their positive index counterpart: $X_{-l}=X_l^*$ for any $l$. 
If we take solutions in the form
\beq
X_l(\vec{r},t) = U_l \; \exp(-\gamma t) \; \exp(i \vec{q}\cdot\vec{r}) \; ,
\eeq
we obtain the eigenvalue equations
\beq
-\gamma U_l +iv \frac{q_x -i q_y}{2} U_{l-1} 
         +iv \frac{q_x +i q_y}{2} U_{l+1} 
= \frac{8 l^2c}{1-4 l^2}  U_l  \; ,
\label{eigen.eq}
\eeq
for $l \in {\mathbb Z}$.
Expanding in powers of the wavenumber $\vec{q}$, the decay rates are given by
\beq
\gamma = \frac{8 l^2c}{4 l^2-1}  - \frac{3(4l^2-1)v^2}{16c} \,
q^2+O(q^4) \; ,
\label{decay.rate}
\eeq
with $l=0,1,2,...$ and $q=\sqrt{\vec{q}^{\, 2}}$.
Figure \ref{fig1} shows the spectrum obtained by solving numerically the
eigenvalue equations (\ref{eigen.eq}) for $c>0$.
We observe that the branch for $l=0$ has a convexity which
is opposite to the one of the other branches for $l=1,2,...$
Moreover, all the branches terminate on the line $\gamma=2c$,
which fixes the maximum value of the wavenumber 
for each branch.  This feature prevents the eigenvalues
to become of the opposite sign hence avoiding instability
in agreement with the stability provided by the $H$-theorem.
For the diffusion sector, the constant $c$ is always positive.
However, the sign of the constant $c$ may change in the reaction sector
and we must treat separately the cases $c>0$ and $c\leq 0$.

\subsection{The case $0<p_0n_r<\frac{n}{2}$}

In this case, we have that $c>0$ in both the diffusion and reaction sectors.

After the relaxation time, the dynamics is dominated 
by the first Fourier components $X_0$, $X_{+1}$, and $X_{-1}$:
\beq
X \simeq X_0 + X_{+1} \; {\rm e}^{+i\varphi} + X_{-1} \; {\rm e}^{-i\varphi} \; .
\eeq
In view of the latter relation, we introduce the
densities
\bega
\rho_+ &\equiv& \rho_A + \rho_B = \int_{-\pi}^{+\pi} f(\vec{r},\varphi) \; d \varphi = 2\pi f_0 \; ,\label{rho+}\\
\rho_- &\equiv& \rho_A - \rho_B = \int_{-\pi}^{+\pi} g(\vec{r},\varphi) \; d \varphi = 2\pi g_0 \; .\label{rho-}
\ega
Thus the zeroth-order expansion of the distribution function $f$ 
can be related to the total density. 
Similarly, we introduce the currents
\bega
\vec{J}_+ &\equiv& \vec{J}_A +  \vec{J}_B = \int_{-\pi}^{+\pi} \vec{v} \; f(\vec{r},\varphi) \; d \varphi  \; ,\\
\vec{J}_- &\equiv& \vec{J}_A - \vec{J}_B =\int_{-\pi}^{+\pi} \vec{v} \; g(\vec{r},\varphi) \; d \varphi  \; .
\ega
The first-order Fourier components $X_{+1}$ and $X_{-1}$
are related to the current. 
As a consequence, we obtain the distribution functions $f$ and $g$ in terms of the corresponding densities and currents
\bega
f  &\simeq& \frac{1}{2\pi}  \left( \rho_+ +\frac{2}{v^2} \vec{v} \cdot \vec{J}_+\right)\; , \label{eq:f}\\
g  &\simeq& \frac{1}{2\pi}  \left( \rho_- +\frac{2}{v^2} \vec{v} \cdot \vec{J}_-\right)\; . \label{eq:g}
\ega
The distribution functions for the species $A$ and $B$ are thus given by
\bega
f_A &\simeq& \frac{1}{2\pi}  \left( \rho_A +\frac{2}{v^2} \vec{v} \cdot \vec{J}_A\right) \; ,\label{eq:f_A}\\
f_B  &\simeq& \frac{1}{2\pi}  \left( \rho_B +\frac{2}{v^2} \vec{v} \cdot \vec{J}_B\right) \; .\label{eq:f_B}
\ega

Equations (\ref{eq:PDEFourier}) for $l=0$ and $l=\pm 1$ then lead to the coupled equations:
\begin{eqnarray}
\partial_t\rho_+ &=& - \vec{\nabla}\cdot \vec{J}_+  \; ,\label{eq.rho+}\\
\partial_t \vec{J}_+&=& - \frac{8av}{3}n \vec{J}_+ -\frac{v^2}{2} \vec{\nabla}\rho_+ \; ,
\end{eqnarray}
and 
\begin{eqnarray}
\partial_t\rho_- &=& - \vec{\nabla}\cdot \vec{J}_- - 4avp_0 n_r \rho_-  \; ,\label{eq.rho-}\\
\partial_t \vec{J}_-&=& - \frac{8av}{3}\left( n- \frac{p_0n_r}{2}\right) \vec{J}_- -\frac{v^2}{2} \vec{\nabla}\rho_-\; .
\end{eqnarray}
We notice that the currents relax on a fast time scale so that we can assume that they quickly adjust
to their value in a quasi-stationary state as 
\begin{eqnarray}
\vec{J}_+&=& - D \vec{\nabla}\rho_+ \; , \qquad \mbox{for} \quad t \gg \frac{3}{8avn} \; ,\label{Fick.D}\\
\vec{J}_-&=& - D^{\rm (r)} \vec{\nabla}\rho_- \; , 
\qquad \mbox{for} \quad t \gg \frac{3}{8av\left(n-\frac{p_0n_r}{2}\right)} \; ,\label{Fick.Dr}
\end{eqnarray}
with the diffusion coefficient
\begin{equation}
D = \frac{3v}{16an} \; ,
\label{diff.coeff}
\end{equation}
and the reactive diffusion coefficient
\begin{equation}
D^{\rm (r)} = \frac{3v}{16a\left(n-\frac{p_0n_r}{2}\right)} \; .
\end{equation}
Equation (\ref{Fick.D}) is the expression of Fick's law for the particles while
Eq. (\ref{Fick.Dr}) is its reactive analogue.
Substituting Eqs. (\ref{Fick.D}) and (\ref{Fick.Dr}) into Eqs. (\ref{eq.rho+})
and (\ref{eq.rho-}) for the densities, we obtain the diffusion equation
\begin{equation}\label{diffusion}
\partial_t \rho_+ = D \nabla^2 \rho_+\; ,
\end{equation}
as well as the reaction-diffusion equation
\begin{equation}\label{react-diff}
\partial_t \rho_- = D^{\rm (r)} \nabla^2 \rho_- - 2 \kappa \rho_- \; ,
\end{equation}
with the reaction rate constant
\begin{equation}
\kappa = 2avp_0n_r \; .
\end{equation}
This result shows that the reaction rate constant is the 
product of the speed $v$ with the cross section $2a$ of the disks, 
multiplied by the density $n_r$ of the catalytic scatterers and
weighted by the probability of reaction $p_0$.  
The two equations (\ref{diffusion}) and (\ref{react-diff}) show the existence
of two slow modes in the system corresponding to the decay rate (\ref{decay.rate}) with $l=0$, namely, the {\it diffusive mode}
of dispersion relation
\begin{equation}
\mbox{diffusive mode:} \qquad \Gamma = D q^2 + O(q^4) \; ,
\end{equation}
and the {\it reactive mode} of dispersion relation
\begin{equation}
\mbox{reactive mode:} \qquad \Gamma = 2\kappa + D^{\rm (r)} q^2 + O(q^4) \; .
\end{equation}

The equations of motion (\ref{diffusion}) and (\ref{react-diff}) 
determines the time evolution of the densities 
$\rhoA = (\rhoplus+\rhominus)/2$ and $\rhoB = (\rhoplus-\rhominus)/2$ 
according to the coupled reaction-diffusion equations:
\begin{mathletters}
\bega
\6_t \rhoA 
    &=& D_{AA} \nabla^2 \rhoA +  D_{AB} \nabla^2 \rho_B
                       -\kappa (\rhoA-\rhoB) \; ,
\label{eq:eq-motA} \\
\6_t \rhoB
    &=&  D_{BA} \nabla^2 \rhoA +  D_{BB} \nabla^2 \rhoB  
                       +\kappa (\rhoA-\rhoB) \; ,
\label{eq:eq-motB}
\ega 
\label{eq:eq-motA-motB}
\end{mathletters} 
where the transport coefficients can be identified as
\begin{eqnarray}
D_{AA} &=& D_{BB} = \frac{D+D^{\rm (r)}}{2} \; ,\\
D_{AB} &=& D_{BA} = \frac{D-D^{\rm (r)}}{2}\; .
\end{eqnarray}
The important conclusion is here that there appears
a phenomenon of cross diffusion which is induced by the reaction.
Indeed, the cross-diffusion coefficient takes the value
\begin{equation}
D_{AB} = D_{BA} = -\frac{3vp_0n_r}{64an\left(n-\frac{p_0n_r}{2}\right)}\; ,
\end{equation}
which vanishes with the reaction probability $p_0$
and the density of catalysts $n_r$.  In the absence of reaction,
the cross-diffusion terms vanish with the reaction term and we
recover two uncoupled diffusion equations for $A$ and $B$ particles.

We notice that the coupled reaction-diffusion equations (\ref{eq:eq-motA})-(\ref{eq:eq-motB}) can be rewritten as
\begin{mathletters}
\bega
\6_t \rhoA 
    &=& -\vec{\nabla}\cdot \vec{J}_A
                       -\kappa (\rhoA-\rhoB) \; ,
\label{eq:eq-JA} \\
\6_t \rhoB
    &=&  -\vec{\nabla}\cdot \vec{J}_B
                           +\kappa (\rhoA-\rhoB) \; ,
\label{eq:eq-JB}
\ega 
\end{mathletters} 
in terms of the currents:
\bega
\vec{J}_A 
&=& 
- \frac{D+D^{\rm (r)}}{2} \vec{\nabla} \rhoA - \frac{D-D^{\rm(r)}}{2} \vec{\nabla} \rhoB \; ,
\label{eq:eqjA}
\\
\vec{J}_B 
&=&
- \frac{D+D^{\rm(r)}}{2} \vec{\nabla} \rhoB - \frac{D-D^{\rm(r)}}{2} \vec{\nabla} \rhoA\; .
\label{eq:eqjB}
\ega

\begin{figure}[htbp]
\centerline{\includegraphics[width=16cm]{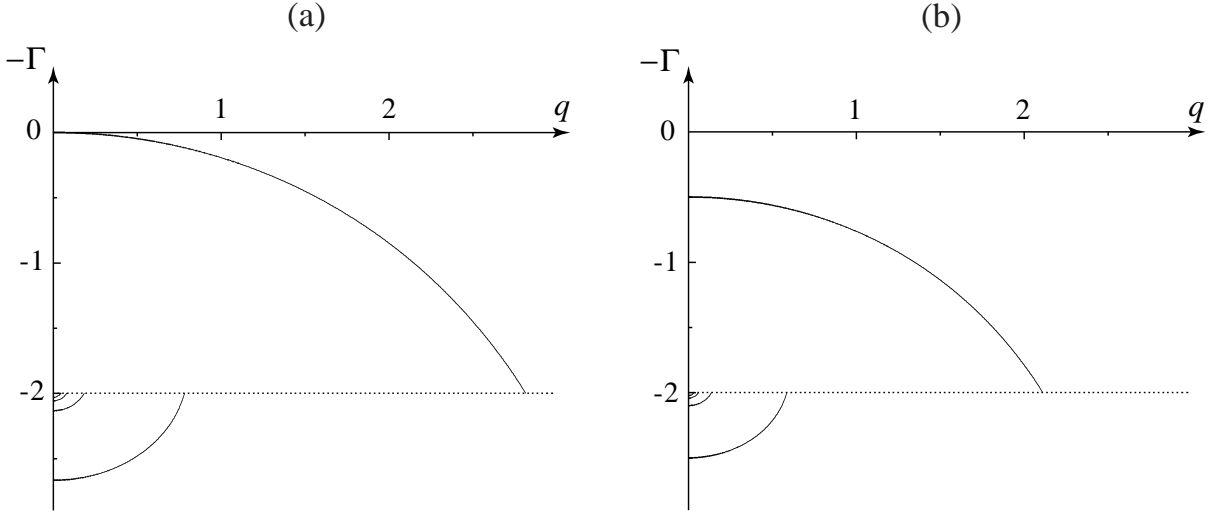}}
\caption{Eigenvalues $-\Gamma$ versus the wavenumber $q=q_x$ for $q_y=0$
in the regime $0<p_0n_r<\frac{n}{2}$ where $c>0$: (a) spectrum of the diffusion sector;
(b) spectrum of the reaction sector. The parameter values are $av=1$, $n=1$, and $p_0n_r=0.125$.}
\label{fig2}
\end{figure}

Besides the diffusive and reactive slowest modes, we also find faster modes
often referred to as kinetic modes.  All these modes exist in both the diffusion sector
ruled by Eq. (\ref{eq:X}) for $X=f$ and in the reaction sector ruled by Eq. (\ref{eq:X})  for $X=h$.  
All these modes are characterized by dispersion
relations which form a whole spectrum.  Figure \ref{fig2} depicts the whole spectra
of the diffusionand reaction sectors
in the case $0<p_0n_r<\frac{n}{2}$.

\subsection{The case $p_0n_r=\frac{n}{2}$}

In this case, we remark that Eq. (\ref{eq:X}) for $X=h$ has a vanishing coefficient 
$c=0$ in the reaction sector so that the equation for $X=h$ is purely advective
\begin{equation}
\partial_t h+\vec{v}\cdot\vec{\nabla} h=0 \; .
\end{equation}
Its solutions are given by $h=h(\vec{r}-\vec{v}t)$ and $g={\rm e}^{-2\kappa t} h(\vec{r}-\vec{v}t)$ obeys
\begin{equation}
\partial_t g+\vec{v}\cdot\vec{\nabla} g= - 2 \kappa g \; .
\end{equation}
In this case, there is no reactive diffusion coefficient which characterizes
the reactive process.

\subsection{The case $\frac{n}{2}<p_0n_r<n$}

As we noticed before, $p_0n_r$ cannot exceed the value $n$ for consistency.
In this case, we have that Eq. (\ref{eq:X}) for $X=f$ still has a coefficient $c>0$ 
in the diffusion sector but Eq. (\ref{eq:X}) for $X=h$ has a negative coefficient $c<0$ in the reaction sector.
Accordingly, the spectrum shown in Fig. \ref{fig1} is upside down in the reaction sector and
the slowest reactive mode is no longer the same as before.

Here, we must consider the decay rate (\ref{decay.rate}) with $l=1$.
The dispersion relation of the reactive mode is now given by
\begin{equation}
\Gamma = 2\kappa' + {D^{\rm (r)}}' q^2 \; ,
\end{equation}
with the new reaction constant
\begin{equation}
\kappa' = \frac{2av}{3} (2n-p_0n_r)\; ,
\end{equation}
and the new reactive diffusion coefficient
\begin{equation}
{D^{\rm (r)}}' = \frac{9v}{16a(2p_0n_r-n)}\; .
\end{equation}
A transition should therefore appear at high concentrations of catalysts.

\begin{figure}[htbp]
\centerline{\includegraphics[width=16cm]{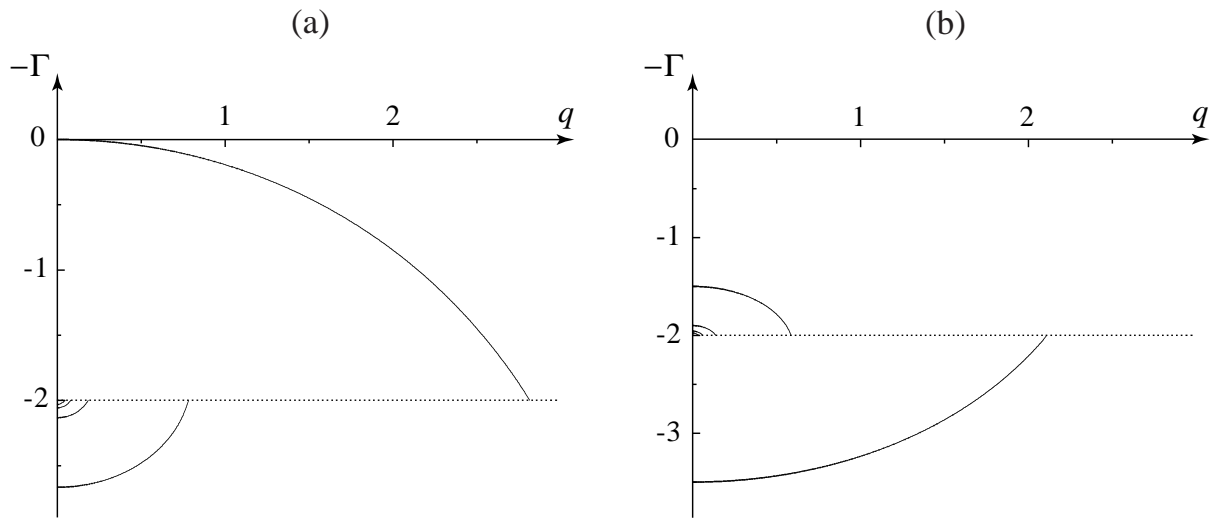}}
\caption{Eigenvalues $-\Gamma$ versus the wavenumber $q=q_x$ for $q_y=0$
in the regime $\frac{n}{2}<p_0n_r<n$ where $c>0$: (a) spectrum of the diffusion sector;
(b) spectrum of the reaction sector. The parameter values are $av=1$, $n=1$, and $p_0n_r=0.875$.}
\label{fig3}
\end{figure}

Figure \ref{fig3} depicts the spectra in the diffusion and reaction sectors
in the case $\frac{n}{2}<p_0n_r<n$.

In the following, we shall focus on the case $0<p_0n_r<\frac{n}{2}$
where the catalysts are dilute enough.

%%******************************************
%%******************************************
\section{The entropy balance} 
\label{sec:entropybalance}

In this section, we derive the equation for the balance of entropy
from the $H$-theorem in two approximations for the entropy density.
In kinetic theory, we have a well-defined expression for the entropy
which guarantees that the entropy production is always non-negative.
However, at the level of the macroscopic description given by the
reaction-diffusion equations (\ref{eq:eq-motA})-(\ref{eq:eq-motB}),
the entropy is given as an approximation of the expression (\ref{eq:s-Boltzmann})
of kinetic theory and we must verify the domain of validity
where the corresponding entropy production is non-negative.

The first approximation we consider is based on the standard phenomenological entropy
defined in irreversible thermodynamics of reaction processes \cite{dGM,KoPr98}.
We show that the corresponding entropy production is non-negative
in a broad range of values for the densities $\rho_A$ and $\rho_B$
including the thermodynamic equilibrium state but there is a small
domain where the entropy production corresponding to this approximation
fails to remain non-negative.

Therefore, we consider a second approximation which includes
extra terms involving the gradient of the densities.
We show that the corresponding entropy production always remains
non-negative.

\subsection{Entropy without gradients}

Supposing that the system is sufficiently dilute, the phenomenological
irreversible thermodynamics supposes that the
entropy density has the following expression: 
\beq
\label{eq:s-thermo}
s=\rhoA \ln\frac{\rho^0}{\rhoA} + \rhoB \ln\frac{\rho^0}{\rhoB}\; .
\eeq
This entropy density is obtained from the entropy (\ref{eq:s-Boltzmann})
of kinetic theory by using the expansions (\ref{eq:f_A}) and (\ref{eq:f_B})
of the distribution functions and by keeping the terms in the densities
themselves and discarding terms in the gradients.
The reference density is thus given by $\rho^0=2\pi f^0$, 
which amounts to suppose the equality 
of the masses of the particles $A$ and $B$. 

The variation of the entropy density $s$ in time is given by 
\beq
\6_t s = -\vec{\nabla} \cdot \vec{J}_s + \sigma_s\; ,
\eeq
as calculated by using the coupled reaction-diffusion equations
(\ref{eq:eq-JA})-(\ref{eq:eq-JB}).
The entropy current density is obtained in terms of the currents (\ref{eq:eqjA})
and (\ref{eq:eqjB}) of particles $A$ and $B$ as
\beq
\label{eq:entropycurrent}
\vec{J}_s = \vec{J}_A \ln \frac{\rho^0}{{\rm e}\rhoA} +\vec{J}_B \ln\frac{\rho^0}{{\rm e}\rhoB}\; ,
\eeq
while the entropy production as
\begin{equation}\label{eq:sigma-thermo}
\sigma_s = \kappa (\rhoA-\rhoB) \ln \frac{\rhoA}{\rhoB}
 - \vec{J}_A\cdot \frac{\vec{\nabla} \rhoA}{\rhoA} 
 - \vec{J}_B\cdot \frac{\vec{\nabla} \rhoB}{\rhoB} \; .
 \end{equation}
Inverting Eqs. (\ref{eq:eqjA}) and (\ref{eq:eqjB}), we obtain the gradients
in terms of the currents as
\begin{eqnarray}
\vec{\nabla}\rho_A &=& - \frac{\vec{J}_A}{D} + \frac{2\kappa}{3v^2} (\vec{J}_A-\vec{J}_B)\label{grad_A}\; , \\
\vec{\nabla}\rho_B &=& - \frac{\vec{J}_B}{D} - \frac{2\kappa}{3v^2} (\vec{J}_A-\vec{J}_B) \label{grad_B}\; .
\end{eqnarray}
Substituting in the entropy production (\ref{eq:sigma-thermo}), we get
\begin{equation}\label{eq:sigma-thermo.bis}
\sigma_s = \kappa (\rhoA-\rhoB) \ln \frac{\rhoA}{\rhoB}
 +\alpha \vec{J}_A^2 + 2 \beta  \vec{J}_A\cdot  \vec{J}_B + \gamma  \vec{J}_B^2 \; ,
 \end{equation}
 with the coefficients
 \begin{eqnarray}
 \alpha &=& \frac{1}{\rho_A}\left( \frac{1}{D} - \frac{2\kappa}{3v^2}\right) \; ,\\
 \beta &=& \frac{\kappa}{3v^2}\left( \frac{1}{\rho_A} + \frac{1}{\rho_B}\right) \; ,\\
 \gamma &=& \frac{1}{\rho_B}\left( \frac{1}{D} - \frac{2\kappa}{3v^2}\right) \; .
 \end{eqnarray}
 
The first term in the entropy production (\ref{eq:sigma-thermo.bis})
is always non-negative because $(x-y)\ln(x/y)\geq 0$ for all positive values of 
the real numbers $x$ and $y$.  On the other hand, the last terms constitute
a quadratic form which is non-negative under the conditions that $\beta^2-\alpha\gamma <0$
with $\alpha>0$ and $\gamma>0$.  For the diffusion coefficient (\ref{diff.coeff}),
we check that
\begin{eqnarray}
\alpha &=&\frac{4a}{3v\rho_A}(4n-p_0n_r) >0 \; , \\
\gamma &=&\frac{4a}{3v\rho_B}(4n-p_0n_r) >0 \; ,
\end{eqnarray}
because of the consistency condition (\ref{consistency}).
Next, the condition $\beta^2-\alpha\gamma <0$ is given by
\begin{equation}\label{eq:xi}
\xi^2 + 2 \left[ 1-2 \left(\frac{2}{y}-1\right)^2\right] \, \xi + 1 < 0 \; ,
\end{equation}
in terms of the density ratio
\begin{equation}\label{ratio.xi}
\xi \equiv \frac{\rho_B}{\rho_A} \; ,
\end{equation}
and the parameter
\begin{equation}\label{param.y}
y \equiv \frac{p_0n_r}{2n} = \frac{4\kappa D}{3v^2} \; .
\end{equation}
The roots $\xi_{\pm}$ of Eq. (\ref{eq:xi}) satisfy $\xi_+ = \frac{1}{\xi_-}$.
If $\xi_+\geq \xi_-$, the entropy production is thus non-negative in the domain
\begin{equation}\label{domain.non-neg}
\xi_+ \rho_A \geq \rho_B \geq \xi_- \rho_A \; .
\end{equation}
It turns out that the sum of the roots
\begin{equation}
\xi_++\xi_- = 2 \left[ 2 \left(\frac{2}{y}-1\right)^2 - 1\right] \; ,
\end{equation}
is always positive in the interval $0<y<\frac{1}{2}$ where the condition 
of consistency (\ref{consistency}) is verified. (The sum $\xi_++\xi_-$ is negative
only in the interval $4-2\sqrt{2}=1.1715...<y<4+2\sqrt{2}=6.8284...$ outside the domain of consistency.)
Since $\xi_+$ and $\xi_-$ have the same sign, they both are positive in the consistency domain
$0<y<\frac{1}{2}$, which means that there exists a domain of the quadrant ($\rho_A \geq 0$,
$\rho_B \geq 0$) where the entropy production (\ref{eq:sigma-thermo.bis}) can be negative.
This domain is composed of $\xi_-\; \rho_A > \rho_B \geq 0$ and $\xi_-\; \rho_B > \rho_A \geq 0$.
Another way of saying it is that the domain (\ref{domain.non-neg}) where
the entropy production is non-negativity is smaller than the quadrant ($\rho_A \geq 0$,
$\rho_B \geq 0$) of all the possible densities $\rho_A$ and $\rho_B$.
The domain of non-negativity extends to
\begin{equation}
1441.9... \, \rho_A \geq \rho_B \geq 0.00069348... \, \rho_A \; , \qquad \mbox{for} \quad y=0.1 \; ,
\end{equation}
to
\begin{equation}
321.99... \, \rho_A \geq \rho_B \geq 0.0031056... \, \rho_A \; , \qquad \mbox{for} \quad y=0.2 \; ,
\end{equation}
and to
\begin{equation}
33.970... \, \rho_A \geq \rho_B \geq 0.029437... \, \rho_A \; , \qquad \mbox{for} \quad y=0.5 \; .
\end{equation}
It is only in the limit $y=0$ without chemical reaction that the domain of non-negativity
coincides with the quadrant ($\rho_A \geq 0$, $\rho_B \geq 0$).
In the presence of the chemical reaction, the situation is unsatisfactory
because the scheme is not consistent with the second law of thermodynamics
even if the domain of negative entropy production is small and only concerns
color densities which are very far from the thermodynamic equilibrium.

\subsection{Entropy with gradients}

To cure the problem reported here above, we choose to expand the
entropy production (\ref{eq:s-Boltzmann}) of kinetic theory to include 
the terms with gradients of the densities.
For this purpose, we substitute the expansions (\ref{eq:f_A}) and (\ref{eq:f_B})
of the distribution functions in the entropy (\ref{eq:s-Boltzmann}) and we truncate
up to the terms which are quadratic in the currents to get
\beq
\label{eq:s-grad}
s=\rhoA \ln\frac{\rho^0}{\rhoA} + \rhoB \ln\frac{\rho^0}{\rhoB} - \frac{1}{v^2} \left( \frac{\vec{J}_A^2}{\rho_A} + \frac{\vec{J}_B^2}{\rho_B}\right) \; ,
\eeq
where the particle currents are given in terms of
the density gradients according to Eqs. (\ref{eq:eqjA}) and (\ref{eq:eqjB}).
Accordingly, the entropy (\ref{eq:s-grad}) is quadratic in the density gradients.

Here again, we calculate
the time variation of the entropy density $s$ by using the coupled reaction-diffusion equations
(\ref{eq:eq-JA})-(\ref{eq:eq-JB}).  The balance equation for entropy is here also
given by
\beq
\6_t s = -\vec{\nabla} \cdot \vec{J}_s + \sigma_s \; ,
\eeq
with an entropy current similar to Eq. (\ref{eq:entropycurrent})
\beq
\vec{J}_s = \vec{J}_A \ln \frac{\rho^0}{{\rm e} \rhoA} +\vec{J}_B \ln\frac{\rho^0}{{\rm e}\rhoB} + O(3) \; ,
\eeq
up to the terms $O(3)$ of third order in the gradients.
However, the entropy production now takes the more complicated form
\begin{equation}\label{eq:entr.prod.grad}
\sigma_s = \kappa (\rhoA-\rhoB) \ln \frac{\rhoA}{\rhoB}
 - \vec{J}_A\cdot \frac{\vec{\nabla} \rhoA}{\rhoA} 
 - \vec{J}_B\cdot \frac{\vec{\nabla} \rhoB}{\rhoB} 
 + \frac{2\kappa}{v^2} (\vec{J}_A-\vec{J}_B)\cdot
 \left( \frac{\vec{J}_A}{\rho_A} - \frac{\vec{J}_B}{\rho_B}\right)
- \frac{\kappa}{v^2} (\rho_A-\rho_B)
\left( \frac{\vec{J}_A^2}{\rho_A^2}- \frac{\vec{J}_B^2}{\rho_B^2}\right)+O(4) \; ,
\end{equation}
where $O(4)$ denotes terms of fourth order in the gradients.
Replacing the gradients by the currents with Eqs. (\ref{grad_A})-(\ref{grad_B}),
we obtain
\begin{equation}\label{eq:entr.prod.grad.bis}
\sigma_s = \kappa (\rhoA-\rhoB) \ln \frac{\rhoA}{\rhoB}
 + \frac{1}{D} \left( \frac{\vec{J}_A^2}{\rho_A} + \frac{\vec{J}_B^2}{\rho_B}\right)
 + \frac{4\kappa}{3v^2} (\vec{J}_A-\vec{J}_B)\cdot
 \left( \frac{\vec{J}_A}{\rho_A} - \frac{\vec{J}_B}{\rho_B}\right)
- \frac{\kappa}{v^2} (\rho_A-\rho_B)
\left( \frac{\vec{J}_A^2}{\rho_A^2} - \frac{\vec{J}_B^2}{\rho_B^2}\right)+O(4) \; ,
\end{equation}
or equivalently
\begin{equation}\label{eq:entr.prod.grad.ter}
\sigma_s = \kappa (\rhoA-\rhoB) \ln \frac{\rhoA}{\rhoB}
 +\alpha \vec{J}_A^2 + 2 \beta  \vec{J}_A\cdot  \vec{J}_B + \gamma  \vec{J}_B^2+O(4) \; ,
 \end{equation}
 with the coefficients
 \begin{eqnarray}
 \alpha &=& \frac{1}{\rho_A}\left( \frac{1}{D} +\frac{\kappa}{3v^2}+ \frac{\kappa\rho_B}{v^2\rho_A}\right) \; ,\\
 \beta &=& -\frac{2\kappa}{3v^2}\left( \frac{1}{\rho_A} + \frac{1}{\rho_B}\right) \; ,\\
 \gamma &=& \frac{1}{\rho_B}\left( \frac{1}{D} +\frac{\kappa}{3v^2}+ \frac{\kappa\rho_A}{v^2\rho_B}\right)\; .
 \end{eqnarray}
 The coefficients $\alpha$ and $\gamma$ are always positive, while the
 condition $\beta^2-\alpha\gamma < 0$ of non-negativity of the quadratic form is here given by
 \begin{equation}\label{eq:xi.grad}
\xi^2 + 2 \frac{(y+2)^2+4}{y(12-y)} \, \xi + 1 > 0 \; ,
\end{equation}
with the ratio $\xi$ defined by Eq. (\ref{ratio.xi}) and the parameter $0<y<\frac{1}{2}$ 
by Eq. (\ref{param.y}).
The opposite inequality is obtained compared to Eq. (\ref{eq:xi}) because the dependence
on the ratio $\xi$ is here more complicated but still simple enough to lead to the quadratic
equation (\ref{eq:xi.grad}).  In the physical domain $0<y<\frac{1}{2}$, the roots $\xi_+$ and
$\xi_-=1/\xi_+$ of Eq. (\ref{eq:xi.grad}) are real and negative because
 \begin{equation}
\xi_++\xi_- = - 2 \frac{(y+2)^2+4}{y(12-y)} < 0 \; .
\end{equation}
As a consequence, the quadratic part of the entropy production (\ref{eq:entr.prod.grad.ter})
is non-negative in the quadrant  of all the physically allowed densities where
$\rho_A \geq 0$ and $\rho_B \geq 0$.
If the gradients of densities are small enough so that
the corrections of fourth order in Eq. (\ref{eq:entr.prod.grad.ter}) are negligible,
the whole entropy production (\ref{eq:entr.prod.grad.ter}) is also non-negative.

We have thus proved that the inclusion of the gradient terms in the
entropy avoids the aforementioned problem and guarantees that
the entropy production remains non-negative for all the values of the color densities if the gradients are
sufficiently small.

\subsection{Interpretation of the gradient terms in the entropy}

The entropy density given by Eq. (\ref{eq:s-grad}) can be expressed
as follows in terms of the gradients of the particle densities by
using Eqs. (\ref{eq:eqjA})-(\ref{eq:eqjB})
\beq
\label{eq:s-grad.grad}
s=\rhoA \ln\frac{\rho^0}{\rhoA} + \rhoB \ln\frac{\rho^0}{\rhoB} 
- \frac{K_{AA}}{2} \left( \vec{\nabla}\rho_A\right)^2
- K_{AB} \vec{\nabla}\rho_A\cdot\vec{\nabla}\rho_B
- \frac{K_{BB}}{2} \left( \vec{\nabla}\rho_B\right)^2 \; ,
\eeq
with the coefficients
\begin{eqnarray}
K_{AA}&=&\frac{\left( D+D^{\rm (r)}\right)^2}{2v^2\rho_A}
+\frac{\left( D-D^{\rm (r)}\right)^2}{2v^2\rho_B} \; , \label{KAA} \\
K_{BB}&=&\frac{\left( D+D^{\rm (r)}\right)^2}{2v^2\rho_B}
+\frac{\left( D-D^{\rm (r)}\right)^2}{2v^2\rho_A} \; ,\label{KBB} \\
K_{AB}&=&\frac{\left(D+D^{\rm (r)}\right)\left(D-D^{\rm (r)}\right)}{2v^2}
\left(\frac{1}{\rho_A}+\frac{1}{\rho_B}\right) \; , \label{KAB}
\end{eqnarray}
which are independent of the velocity $v$.

The gradient terms are of the same kind as those appearing in
the Ginzburg-Landau free energy.  Here, they appear in the entropy with the opposite sign in agreement with the required thermodynamic
stability of the equilibrium state \cite{PeFi90,WaSeWh93}.
Indeed, the entropy must be maximal in a stable equilibrium state
which is the case since the quadratic form in Eq. (\ref{eq:s-grad})
or (\ref{eq:s-grad.grad}) is negative.  Accordingly, the entropy reaches a maximum at the equilibrium state where the gradients and the currents vanish.

The gradient terms are responsible for statistical correlations between
the particles.  Indeed, the entropy density (\ref{eq:s-grad.grad}) can be
used to define the entropy functional
\beq
S[\rho_A,\rho_B]=\int s\;  d\vec{r} \; ,
\eeq
and the probability distribution for statistical average given by
the functional integrals
\beq
\langle {\cal O} \rangle_{\rm eq} = \frac{\int{\cal D}\rho_A{\cal D}\rho_B \; {\cal O}\;\exp\left( \frac{1}{k_{\rm B}}S[\rho_A,\rho_B]\right)}
{\int{\cal D}\rho_A{\cal D}\rho_B \; \exp\left( \frac{1}{k_{\rm B}}S[\rho_A,\rho_B]\right)} \; ,
\label{stat}
\eeq
for an observable $\cal O$.  This allows us to calculate
the correlation functions of the particle densities
at the thermodynamic equilibrium.  We consider the correlation
functions of the densities (\ref{rho+})-(\ref{rho-}):
\begin{eqnarray}
C_{++}(\vec{r})&\equiv& \langle \rho_+(\vec{r})\rho_+(0)\rangle_{\rm eq}-\langle \rho_+\rangle_{\rm eq}^2 \; ,\\
C_{--}(\vec{r})&\equiv& \langle \rho_-(\vec{r})\rho_-(0)\rangle_{\rm eq}-\langle \rho_-\rangle_{\rm eq}^2 \; , \\
C_{+-}(\vec{r})&\equiv& \langle \rho_+(\vec{r})\rho_-(0)\rangle_{\rm eq}-\langle \rho_+\rangle_{\rm eq}\, \langle \rho_-\rangle_{\rm eq} \; .
\end{eqnarray}
Partial differential equations can be obtained for these correlation
functions by the variational principle $\delta S=0$ based on the entropy functional 
(\ref{eq:s-grad.grad}) evaluated around the thermodynamic equilibrium
\begin{equation}
s \simeq s_{\rm eq} - \frac{(\rho_+-\rho_{\rm eq})^2+\rho_-^2}{2\, \rho_{\rm eq}} - \frac{K_{++,{\rm eq}}}{2}\left(\vec{\nabla}\rho_+\right)^2
- \frac{K_{--,{\rm eq}}}{2}\left(\vec{\nabla}\rho_-\right)^2 \; ,
\label{eq.entr}
\end{equation}
where the equilibrium values of the coefficients are given by $\langle \rho_+\rangle_{\rm eq}=\rho_{\rm eq}=\frac{2}{\rm e}\rho^0$, $\langle \rho_-\rangle_{\rm eq}=0$, $s_{\rm eq}=\rho_{\rm eq}$, and
\begin{eqnarray}
K_{++,{\rm eq}} &=& \frac{1}{4} \left(K_{AA}+K_{BB}+2K_{AB}\right)_{\rm eq} = \frac{2D^2}{v^2\rho_{\rm eq}} \; , \\
K_{--,{\rm eq}} &=& \frac{1}{4} \left(K_{AA}+K_{BB}-2K_{AB}\right)_{\rm eq} = \frac{2D^{\rm (r)\, 2}}{v^2\rho_{\rm eq}} \; , \\
K_{+-,{\rm eq}} &=& \frac{1}{4} \left(K_{AA}-K_{BB}\right)_{\rm eq} = 0 \; ,
\end{eqnarray}
so that there is no cross term in the gradients of $\rho_{\pm}$ in Eq. (\ref{eq.entr}).
The fluctuations described by Eq. (\ref{stat}) are Gaussian around the equilibrium since the entropy density (\ref{eq.entr}) is quadratic. As a consequence, the correlation functions obey the equations
\begin{eqnarray}
& &\left(\nabla^2-\frac{1}{\ell_+^2}\right)C_{++}(\vec{r})=0 \; , \\
& &\left(\nabla^2-\frac{1}{\ell_-^2}\right)C_{--}(\vec{r})=0 \; ,
\end{eqnarray}
for $\vec{r}\neq 0$ and
with the correlation lengths
\begin{eqnarray}
\ell_+ &\equiv& \sqrt{\rho_{\rm eq}K_{++,{\rm eq}}} = \frac{3\sqrt{2}}{16an} \; ,\\
\ell_- &\equiv& \sqrt{\rho_{\rm eq}K_{--,{\rm eq}}} = \frac{3\sqrt{2}}{16a\left(n-\frac{p_0n_r}{2}\right)}\; ,
\end{eqnarray}
and $C_{+-}(\vec{r})=0$.
These correlation lengths are of the order of the mean free path
of the particles between the scatterers of radius $a$ and density $n$.
The correlation functions are given by Bessel functions of zeroth order
and they behave at long distance as
\beq
C_{\pm\pm}(\vec{r}) \sim \frac{1}{\sqrt{r}} \; \exp\left(-\frac{r}{\ell_{\pm}}\right) \; ,
\eeq
with $r=\sqrt{\vec{r}^{\, 2}}$.

%%*************************************************************
%%*************************************************************
\section{Conclusions}
\label{sec:conclusions}  

In this paper, we have studied a reactive random Lorentz gas in which a point particle
carrying a color $A$ or $B$ (or a spin one-half) bounces among randomly
distributed disk scatterers.  Some of these disks are catalytic in the sense that
the reaction $A\rightleftharpoons B$ occurs with a given probability $p_0$
upon collision on these catalytic disks. In the case of a particle with a spin,
the catalytic disks correspond to impurities flipping the spin.
Under dilute-gas conditions, the time evolution of the distribution functions
of finding the particle with a given color or spin at some position $\vec{r}$
with some velocity $\vec{v}$ are ruled by two coupled Boltzmann-Lorentz kinetic equations
which satisfy a $H$-theorem.  The $H$-quantity defines the entropy at the kinetic level of
description.

The time evolution separates into a diffusion sector for the total distribution function 
and a reaction sector for the difference of the distribution functions of the colors.
The spectrum of eigenmodes of both sectors can be constructed in detail.  
The diffusive (resp. reactive) modes are the slowest modes 
among all the modes of the diffusion (resp. reaction) sector,
which provides us with the diffusion coefficient of the diffusive mode,
as well as the reaction rate constant and the reactive diffusion coefficient
of the reactive mode.  This analysis and the knowledge of these coefficients
allow us to obtain the macroscopic reaction-diffusion equations.
These equations present cross-diffusion terms which are induced by
the reaction if the reaction probability $p_0$ is not vanishing,
as in the previous studies of the reactive periodic Lorentz gas
and multibaker models \cite{GaKl98,Ga99,ClGa00,ClGa02,GaCl02}.
In the reactive random Lorentz gas, our analysis shows that
a transition happens in the reaction sector between a regime 
at low concentrations of catalytic disks and reaction probabilities and another one at
high concentrations and probabilities.

Using the derivation of the macroscopic reaction-diffusion equations, 
we have studied the problem of the entropy production on the basis of the entropy
defined in kinetic theory in terms of the distribution functions
and the associated $H$-theorem.  The entropy of kinetic theory can be 
expanded in powers of the gradients of the densities of both colors $A$ and $B$.
At the lowest order of this expansion, the entropy density is simply a
function of the color densities themselves and coincides with the expression of
the phenomenological nonequilibrium thermodynamics \cite{dGM,KoPr98}.
The balance equation of this entropy without gradient
can be derived from the macroscopic reaction-diffusion equations.
Because of the cross-diffusion effects induced by the reaction, the resulting
entropy production may become negative for extreme values of the ratio
between the color densities.  This is certainly a problem of principle
for the phenomenological approach.  In order to solve this problem,
we have considered the entropy density at the next order of the
expansion in the gradients of color densities.  At this next order,
the entropy density contains terms which are quadratic in the gradients
beside the contribution of the phenomenological entropy.
The balance equation of this entropy with gradients turns out to have
an entropy production which is non-negative for all the
color densities and for small enough gradients of color densities,
in consistency with the second law of thermodynamics.

The entropy density with gradients is interpreted as the entropic
version of the Ginzburg-Landau free energy.  The addition of gradient terms
are shown to be responsible for statistical correlations in the densities
of the colors $A$ and $B$ over spatial scales of the order of the
mean free path of the particle.  The consideration of these spatial
correlations appears to be necessary to get a non-negative entropy production
in the presence of chemically induced cross diffusion.
The inclusion of these gradients in the entropy 
is justified by the kinetic theory and by the consistency so obtained
with the second law of thermodynamics.
Such an entropy functional with gradients is also justifed by analogy with the
Ginzburg-Landau free energy \cite{PeFi90,WaSeWh93}.
The inclusion of gradients in the entropy requires a departure from
the classical Onsager-Prigogine nonequilibrium thermodynamics \cite{dGM,KoPr98}.
This classical nonequilibrium thermodynamics neglects the possible
interplay between the diffusion and the reaction, such as
the cross diffusion induced by the chemical reaction.  This effect
is expected in systems with a high reaction
probability and high concentrations of reactants with respect to the inerts
species which do not participate to the reaction.
We think that the present workk clearly shows that this
chemically induced cross diffusion is compatible with
kinetic theory and the second law of thermodynamics and should be
an experimentally observable effect.

%%*****************************************************************
%%*****************************************************************

\vspace{0.3cm}

{\bf Acknowledgments.} 
The authors thank Professor G.~Nicolis for support and encouragement in
this research. L.M. is supported through a European Community Marie Curie Fellowship 
-- contract No. HPMF-CT-2002-01511.  
This research is financially supported by the ``Communaut\'e fran\c caise de Belgique"
(``Actions de Recherche Concert\'ees", contract No. 04/09-312),
the National Fund for Scientific Research (F.~N.~R.~S. Belgium), 
the F. R. F. C. (contract No. 2.4577.04.), and the U.L.B..

%%*****************************************************************
%%*****************************************************************


\begin{thebibliography}{xx}

\bibitem{EvHobooks}
   D.J.\ Evans and  G.P.\ Morriss,
   {\em Statistical Mechanics of Nonequilibrium Liquids\/}
   (Academic Press, London, 1990);

\bibitem{GaNi90}
P.~Gaspard and G.~Nicolis, Phys.\ Rev.\ Lett.\ {\bf 65}, 1693 (1990).  

\bibitem{Ga92}
  P.~Gaspard, J.~Stat.~Phys., {\bf 68}, 673 (1992).

\bibitem{ChEyLeSi93}
   N.I.\ Chernov, G.L.\ Eyink, J.L.\ Lebowitz, and Ya.G.\ Sinai,
   Phys.\ Rev.\ Lett.\ {\bf 70}, 2209 (1993);
   Comm.\ Math.\ Phys.\ {\bf 154}, 569 (1993).

\bibitem{Ru96}
   D.~Ruelle, J.\ Stat.\ Phys.\, {\bf 85}, 1 (1996); {\bf 86}, 935 (1997).

\bibitem{Gabook}
P.~Gaspard, {\em Chaos, Scattering and Statistical Mechanics}, 
(Cambridge University Press, Cambridge, 1998).

\bibitem{Dobook}
J.R.~Dorfman, {\em An Introduction to Chaos in Nonequilibrium Statistical 
Mechanics}, (Cambridge University Press, Cambridge 1999).

\bibitem{VoTeBr98}
  T.~T\'el, J.~Vollmer, and W.~Breymann, 
  Phys.\ Rev.\ E, {\bf 58}, 1672 (1998).

\bibitem{TeVo00}
T.~T\'el and J.~Vollmer, {\em Entropy balance, Multibaker Maps, and the 
Dynamics of the Lorentz gas}, in: D. Sz\'asz, Editor, 
{\em Hard Ball Systems and the Lorentz Gas} (Springer-Verlag, Berlin, 
2000) pp.\ 367-418. 

\bibitem{MaTeVo00}
L.~M\'aty\'as, T.~T\'el, and J.~Vollmer, 
Phys.\ Rev.\ E {\bf 62}, 349 (2000).

\bibitem{TaGa00}
S.~Tasaki and P.~Gaspard, 
J.\ Stat.\ Phys.\ {\bf 101}, 125 (2000). 

\bibitem{TeVoMa01}
T.~T\'el, J.~Vollmer, and L.~M\'aty\'as, 
Europhys.\ Lett.\ {\bf 53}, 458 (2001). 

\bibitem{GiDo99}
T.~Gilbert and J.R.~Dorfman, 
J. \ Stat. \ Phys. \ {\bf 96}, 225 (1999).

\bibitem{GiDoGa00}
T.~Gilbert, J.R.~Dorfman, and P.~Gaspard, 
Phys. \ Rev. \ Lett. \ {\bf 85}, 1606 (2000).

\bibitem{DoGaGi02}
J.R.~Dorfman, P.~Gaspard, and T.~Gilbert, 
Phys. \ Rev. \ E {\bf 85}, 026110 (2002).

\bibitem{GaNiDo02}
P.~Gaspard, G.~Nicolis, and J.R.~Dorfman, 
Physica A {\bf 323}, 294 (2003).

\bibitem{ViGa03a}
S. Viscardy and P. Gaspard,
Phys. Rev. E {\bf 68}, 041204 (2003).

\bibitem{ViGa03b}
S. Viscardy and P. Gaspard,
Phys. Rev. E {\bf 68}, 041205 (2003).

\bibitem{NiPr77} G. Nicolis and I. Prigogine,
{\em Self-organization in nonequilibrium systems}
(Wiley, New York, 1977).

\bibitem{GaKl98} P. Gaspard and R.
Klages, Chaos {\bf 8}, 409 (1998).

\bibitem{Ga99} P. Gaspard, Physica A {\bf 263}, 315 (1999).


\bibitem{footnote1}
Originally the random Lorentz gas was introduced 
in a work of Lorentz \cite{Lobook05} as a model of electric conduction. 

\bibitem{Lobook05} 
H.\ A.\ Lorentz, Proc.\ Roy.\ Acad.\ Amst.\ {\bf 7}, 438, 585, 684 (1905). 

\bibitem{ClGa00}
I.~Claus and P.~Gaspard, J.\ Stat.\ Phys.\ {\bf 101}, 161 (2000). 

\bibitem{ClGa02}
I.~Claus, and P.~Gaspard, Physica D {\bf 168-169}, 266 (2002). 

\bibitem{GaCl02} P. Gaspard and I. Claus, Phil. Trans.
Roy. Soc. Lond. A {\bf 360}, 303 (2002).

\bibitem{dGM}
S.~de Groot and P.~Mazur, {\em Non-equilibrium thermodynamics}
(North-Holland, Amsterdam, 1962; reprinted by Dover Publ.~Co., 
New York, 1984). 

\bibitem{KoPr98} D. Kondepudi and I. Prigogine,
{\em Modern Thermodynamics} (Wiley, New York, 1998).

\bibitem{Be82}
H.\ van Beijeren, Rev.\ Mod.\ Phys.\ {\bf 54}, 195 (1982). 

\bibitem{BoBuSi83}
C.~Boldrighini, L.~A.~Bunimovich, and Ya.G.~Sinai, J.~Stat.~Phys. 
{\bf 32}, 477 (1983).  

\bibitem{PeFi90}
O.~Penrose and P.C.~Fife, Physica D {\bf 43}, 44 (1990).  

\bibitem{WaSeWh93} 
S.-L.~Wang, R.F.~Sekerka, A.A.~Wheeler, B.T.~Murray, 
S.R.~Coriell, R.J.~Braun, and G.B.~McFadden,  Physica D {\bf 69}, 
189 (1993).   

\end{thebibliography}
\end{document}